\documentstyle[11pt]{article}

\newcommand{\be}{\begin{equation}}
\newcommand{\ee}{\end{equation}}
\newcommand{\bea}{\begin{array}}
\newcommand{\ea}{\end{array}}
\newcommand{\beqa}{\begin{eqnarray}}
\newcommand{\eeqa}{\end{eqnarray}}
\newcommand{\bean}{\begin{eqnarray*}}
\newcommand{\eean}{\end{eqnarray*}}
\newcommand{\eqn}[1]{(\ref{#1})}

\newcommand{\del}{\partial}

\def\up#1{\leavevmode \raise.16ex\hbox{#1}}

\newcommand{\journal}[4]{{\sl #1 }{\bf #2} \up(19#3\up) #4}

\newcommand{\gapproxeq}{\lower .7ex\hbox{$\;\stackrel{\textstyle >}{\sim}\;$}}
\newcommand{\lapproxeq}{\lower .7ex\hbox{$\;\stackrel{\textstyle <}{\sim}\;$}}

\newcounter{appendice}

\def\thebibliography#1{{\bf REFERENCES\markboth
 {REFERENCES}{REFERENCES}}\list
 {[\arabic{enumi}]}{\settowidth\labelwidth{[#1]}\leftmargin\labelwidth
 \advance\leftmargin\labelsep
 \usecounter{enumi}}
 \def\newblock{\hskip .11em plus .33em minus -.07em}
 \sloppy
 \sfcode`\.=1000\relax}

\begin{document}

%\title{\hfill {\small{
%{\textstyle hep-th/9707201}
%}} \\
%\vskip=1truecm
%The Quantum Poincar\'e group as Hidden Symmetry of General Relativity}
%\author{G. Bimonte$^{a}$, R. Musto$^{a}$, A. Stern$^{b}$ and
%P. Vitale $^{a}$\thanks{Contribution to the Proceedings of {\it Quantum 
%Groups, Deformations and Contractions} Istanbul 1997}}
%\maketitle
%\thispagestyle{empty}

\title{\hfill $\mbox{\small{
${\rm\textstyle hep-th/9707201\quad\quad}
$}}$ \\[1truecm]
The Quantum Poincar\'e group as Hidden Symmetry of General Relativity}
\author{G. Bimonte$^{a}$, R. Musto$^{a}$, A. Stern$^{b}$ and
P. Vitale $^{a}$
\thanks{Contribution to the Proceedings of {\it Quantum 
Groups, Deformations and Contractions} Istanbul 1997}}
\maketitle
\thispagestyle{empty}
%%%%%%%%%%%%%%%%%%%%%%%%%%%%%%%%%%%%%%%%%%%%%%%%%%%

\begin{center}
{\it a)  Dipartimento di Scienze Fisiche, Universit\`a di Napoli,\\
Mostra d'Oltremare, Pad.19, I-80125, Napoli, Italy; \\
INFN, Sezione di Napoli, Napoli, ITALY.\\
\small e-mail: \tt bimonte,musto,vitale@napoli.infn.it } \\
{\it b) Department of Physics, University of Alabama,\\
Tuscaloosa, AL 35487, USA.\\
\small e-mail: \tt astern@ua1vm.ua.edu }\\
\end{center}

\begin{abstract}
Using the tools of q--differential calculus and quantum Lie algebras
associated to quantum groups, we find a one--parameter family of q-gauge
theories associated to the quantum group $ISO_q(3,1)$. Although the
gauge fields, that is the spin--connection and the vierbeins are
non--commuting objects depending on a deformation parameter, $q$, it is
possible to construct out of them a metric theory which is insensitive
to the deformation. The Christoffel symbols and the Riemann tensor are
ordinary commuting objects. Hence it is argued that torsionless
Einstein's General Relativity is the common invariant sector of a
one--parameter family of deformed gauge theories. 
\end{abstract} 

\section*{Introduction}
The description of general relativity as a gauge theory is a well
established result in 2+1 dimensions \cite{witten}, where in particular
it turns out to be a Chern Simons theory associated to the Poincar\'e
group $ISO(2,1)$. This equivalence allows for its extension as a q-gauge
theory \cite{cast1} once a deformed Chern--Simons theory is available.
This result has been achieved in \cite{noiPL} where a deformed
Chern--Simons action has been found for multi--parametric, minimally
deformed, quantum groups. In \cite{noi2+1} such a theory has been
specialized to the case of the quantum Poincar\'e group $ISO_q(2,1)$
\cite{cast2}, yielding  a description of 2+1 gravity as a q--gauge
theory. The usual description in terms of dreibeins and
spin--connections associated to the Poincar\'e group, $ISO(2,1)$, is
replaced by deformed dreibeins and spin--connections associated to
$ISO_q(2,1)$, which obey nontrivial braiding relations, the limit q=1
corresponding to the usual, undeformed theory. This yields  a
one--parameter family of Lagrangian and Hamiltonian formalisms for 2+1
gravity (the parameter being the deformation parameter). The metric
tensor, which we need to recover Einstein's theory, is constructed as in
the undeformed case as a suitable bilinear in the dreibeins. Then, the
one--parameter family of Lagrangian descriptions (including the case
q=1), has a common metric sector, that is to say, all the
theories of the family are {\it equivalent} from the point of 
view of dynamics. To be more specific, though the underlying connection
components are non--commuting objects and parameter dependent, the
components of the metric tensor $g_{\mu\nu}$ commute among themselves 
and the field 
equations that they satisfy are formally identical to those of the 
ordinary theory. Remarkably, the
relevant fields of the metric theory, such as Christoffel symbols and
the Riemann tensor, turn out to be ordinary commuting fields having the 
standard form in term of the deformed metric.
This situation is somewhat analogous to  the case
of classical fermionic field theories where commuting   Dirac currents 
are constructed as bilinears in non--commuting fields. 

Although the classical dynamics is equivalent to the usual one, that is
not the case for the Hamiltonian formalism. The one parameter family of
canonical formalisms associated to the deformed symmetry yields
inequivalent theories. This feature should become significant when
quantizing such theories. However, it must be mentioned that the problem
of quantization for systems exhibiting q--symmetry is quite delicate and
we do not know, at the moment, how to solve it. 

In this contribution we will deal with the physically interesting 3+1
dimensional case. The $2+1$ case described above may be regarded as a
toy model to be used as a guide. We will see that the final result, that
is the existence of alternative descriptions for General Relativity
relying on the quantum Poincar\'e group is preserved in the $3+1$
dimensional case \cite{noi3+1}. We will find again a hidden quantum
group structure in General Relativity, though this is achieved in a very
non trivial manner. As it is well known, the major difference among the
two theories is that 2+1 gravity may be described by a CS action,
namely, as a gauge theory, while 3+1 gravity cannot be formulated
as a topological theory. It can be described in terms of 
gauge potentials for the Poincar\'e group $ISO(3,1)$, but the action only
exhibits an invariance under the local Lorentz subgroup. Only when we
impose the torsion to be zero, is the action invariant with respect to
the whole Poincar\'e group. It is torsionless General Relativity which
we will deal with and we will show that there exists a whole
one--parameter family of q--gauge theories associated to the
q--Poincar\'e group $ISO_q(3,1)$, all having the same metric sector in
common, exactly as for the 2+1 dimensional case. Again, the Christoffel
symbols and the Riemann tensor, constructed out of non--commuting
connection components, are given by the usual expression in terms of the
metric tensor, and they commute with {\it all} fields of the theory,
hence being ordinary objects. 

It should be stressed that the equivalence not only holds for the pure 
gravity case, but it also holds in the presence of matter, provided there 
are no sources for torsion. (In fact it is only a non--zero torsion 
which distinguishes the different classical theories from one another, 
each one coupling to a different kind of ``exotic'' matter.) 
We finally mention that a one--parameter family of Hamiltonian 
formalisms has been developed for the 3+1 dimensional case too. We will 
not in this paper be concerned with the canonical formalism. In the 2+1 
dimensional case we refer the reader to \cite{noi2+1} while for the 3+1 
dimensional case a detailed analysis may be found in \cite{noi3+1}.

In section {\bf 1} we briefly review the well known formulation of
General Relativity as a gauge theory. In section {\bf 2} we first
describe the structure of $q$--deformed gauge theories mainly along the
line of Refs. \cite{cast1, noiPL}, then we specialize to the
$q$--Poincar\'e group and show that the gauge formulation of Einstein's
General Relativity may be generalized to a one--parameter family of
deformed gauge theories exhibiting local invariance with respect to the
quantum Lie algebra associated to the quantum group $ISO_q(3,1)$. We
conclude with brief final remarks. 

\section{General Relativity as a Gauge Theory}
Before discussing the q-Poincar\'e theory it is useful to briefly review
the description of general relativity as a Poincar\'e group gauge
theory, so that the deformation procedure will be better understood. The
Poincar\'e group $ISO(3,1)$ may be parameterized by a Lorentz matrix
${\ell}_{ab}$ and Lorentz vector $z_a$, $a,b= 1,...4$.  The former
satisfies the  constraints  of the connected Lorentz group 
\be
\ell_{ab}{ \ell_c}^b=
\ell_{ba} {\ell^b}_c= \eta_{ac}~, ~~
 \det \parallel \ell_{ab} \parallel =    1~,
\label {dolm}
\ee
where $\eta_{ab}$ is the Lorentz metric. The associated Lie algebra is
spanned by ten generators $T_i$, which we split into $M_{ab}$ (the
Lorentz generators) and $P_a$ (the translation generators),  satisfying 
\beqa
[M_{ab}, M_{cd}] &=& \eta_{ac} M_{bd} - \eta_{bc} M_{ad} +
\eta_{bd} M_{ac} - \eta_{ad} M_{bc} \cr
 [M_{ab}, P_c] &=& -(\eta_{bc} P_{a}
- \eta_{ac} P_{b} ) \cr
[P_a, P_b] &=& 0~. \label{lie}
 \eeqa
On the space of {\it commutative} functions generated by $\ell_{ab} $
and $z_a$ (the Lie group manifold), one can construct the usual
differential geometry of Maurer-Cartan forms on the Poincar\'e group. 
The Maurer-Cartan form can  be elevated to connection one form
$A(x)=A_\mu^a T_a dx^\mu$, which may be split into spin connection
$~\omega(x) =\omega_\mu^{ab}(x)M_{ab} dx^\mu ~ $ and vierbein one form
$~e(x)=e_\mu^a(x)P_a dx^\mu~.~ $ The derivative 
\be
{\cal D}_\mu =\del_\mu + A_\mu(x) \equiv \del_\mu + A^i_\mu(x) T_i~,
\ee
is covariant with respect to infinitesimal gauge transformations of the
form 
\beqa
\delta \omega^{ab}_\mu & =& \del_\mu\tau^{ab} + \omega^{ac}_{\mu } ~
{\tau_c}^b - \omega^{bc}_{\mu } ~{\tau_c}^{a} ~,\cr
\delta e^c_\mu & = & \del_\mu \rho^c +
\omega^{cb}_{\mu} ~\rho_b - {\tau^c}_b~ e_{\mu }^b~,  \label{gauge}
\end{eqnarray} 
where the gauge parameters $\tau^{ab}=- \tau^{ba}$ and $\rho^a$ are
associated with Lorentz transformations and translations, respectively. 

The  field strength tensor, $F$,  given  by
\be
F = dA +A^2 \label{F}  ~,
\ee
may be split into  the Lorentz curvature ${\cal R}={\cal
R}_{\mu\nu}^{ab} (x)M_{ab} dx^\mu\wedge dx^\nu~$ and the torsion ${\cal
T}={1\over 2}{\cal T}_{\mu\nu}^{a} (x)P_a dx^\mu\wedge dx^\nu ~,$ where 
\beqa
{\cal R}_{\mu\nu}^{ab} &=& \del_\mu \omega_\nu^{ab} -\del_\nu
\omega_\mu^{ab}- [\omega_\mu, \omega_\nu]^{ab} \cr
{\cal T}_{\mu\nu}^{a} &=& \del_\mu e_\nu^{a} -\del_\nu
e_\mu^{a}- [\omega_\mu, e_\nu]^{a}~. \label{RT}
\eeqa

The dynamics of the theory is determined by a locally Lorentz invariant 
action
\be
S= {1\over 4} \int_M \epsilon_{abcd} {\cal R}^{ab} \wedge 
e^c\wedge e^d
\ee
which can be put into the Palatini form
\be
S= {1\over 2} \int d^4 x~{\tt e}~ e^\mu_a e^\nu_b {\cal R}_{\mu\nu}^{ab}
~.\label{S}
\ee 
Here ${\tt e}\equiv$ det $e_\mu^a$ and  $e^{\mu}_a$ denotes the inverse
of the vierbein fields. By taking variations with respect to the vierbeins
and spin--connections we obtain the equations of motion: 
\be
e^\nu_b{\cal R}^{ab}_{\mu\nu} - {1\over 2}e_\mu^a(e^\rho_c e^\nu_b {\cal
R}^{cb}_{\rho\nu})=0 ~,\qquad
{\cal T}^a_{\mu\nu}=0   ~. \label{eqmot}
\ee

To recover Einstein's theory we have to re-express the Palatini action in
terms of the space--time metric and scalar curvature. The space--time
metric is introduced as a bilinear in the vierbeins
\be
{\tt g}_{\mu\nu}=\eta_{ab} e^a_\mu e^b_\nu~,   \label{stm}
\ee
which is symmetric and invariant with respect to local Lorentz
transformations. The Christoffel symbols $\Gamma _{\mu \nu }^\sigma $ 
are then defined by demanding the covariant derivative of the vierbeins 
 to vanish 
\be
0={\tt D}_\mu e_\nu ^b=   \partial _\mu e_\nu ^b+\omega _\mu ^{bc}e_{\nu c}
-\Gamma _{\mu \nu }^\sigma e_\sigma ^b~.\label{covd}
\ee
By multiplying this expression by $\eta_{ab} e^a_\rho$, summing  over
the $b$ index, and symmetrizing  with respect to the space-time indices 
$\nu $ and   $\rho $, we eliminate the spin--connection, getting 
\begin{equation}
0=\eta _{ab}[e_\rho ^a\partial _\mu e_\nu ^b+e_\nu ^a\partial
_\mu e_\rho ^b-e_\rho ^ae_\sigma ^b\Gamma _{\mu \nu }^\sigma -e_\nu
^ae_\sigma ^b\Gamma _{\mu \rho }^\sigma ]~.  \label{eqn13}
\end{equation}
Adding to this the equation obtained by switching $\mu $ and $\nu $, and
subtracting the equation obtained by replacing  indices ($\mu ,\nu ,\rho
)$   by ($\rho ,\mu ,\nu )$, we finally obtain the Christoffel symbols
in the form 
$$
2\eta _{ab}e_\rho ^ae_\sigma ^b\Gamma _{\mu \nu }^\sigma
=\eta _{ab}[e_\rho ^a(\partial _\mu e_\nu ^b+\partial _\nu
e_\mu ^b)+e_\nu ^a(\partial _\mu e_\rho ^b-\partial _\rho e_\mu ^b)
$$
\be
+e_\mu^a(\partial _\nu e_\rho ^b-\partial _\rho e_\nu ^b)]
\ee
or
\be
2{\tt g}_{\rho \sigma }\Gamma _{\mu \nu }^\sigma =\partial _\mu {\tt g}
_{\rho \nu }+\partial _\nu {\tt g}_{\rho \mu }-\partial _\rho {\tt g}_{\nu
\mu }~.  \label{chris}
\ee
The Riemann tensor is then defined as
\beqa
{\tt R}_{\mu \nu \rho}^{~~~~\sigma} v_{\sigma} &=&({\tt D}_{\mu} {\tt
D}_{\nu}-{\tt D}_{\nu} {\tt D}_{\mu} ) v_{\rho} =\cr
 &-&[\del_\mu \Gamma_{\nu\rho}^\sigma -
\del_\nu \Gamma_{\mu\rho}^\sigma + \Gamma_{\mu\tau}^\sigma
\Gamma_{\nu\rho}^\tau - \Gamma_{\nu\tau}^\sigma
\Gamma_{\mu\rho}^\tau] v_\sigma \label{Rtensor}      ~,
\eeqa
where $v_{\mu}$ is an arbitrary covector. Its relation with the Lorentz
curvature is then given by: 
\be
{\tt R}_{\mu \nu \rho}^{~~~~\tau}=- {\cal  R}_{\mu \nu ~~ b}^{~~~a}
e^b_\rho e^\tau_a~.   \label{eqn17}
\ee
Thus it can be checked that the action \eqn{S} takes the usual
Einstein--Hilbert form
\be
S={1\over 2} \int d^4x~ \sqrt{-{\tt g}}~ {\tt R}    \label{EHa}
\ee
where ${\tt g}\equiv$ det ${\tt g}_{\mu\nu}$ and ${\tt R}$ is the scalar
 curvature ${\tt R}={ {\tt R}_{\nu\mu\rho}}^\mu {\tt g}^{\nu\rho}~ $.
The equations of motion now read 
\be
{\tt R}_{\mu\nu}-{1\over 2}{\tt g}_{\mu\nu}{\tt R} = 0 \label{eqm}~,~~~
{\cal T}_{\mu\nu}^\rho = 0
\ee
where $ {\cal T}_{\mu\nu}^a=e^a_\rho {\cal T}_{\mu\nu}^\rho~$.

This discussion can be easily extended to take into account spinless
matter. The matter action  has to be added to \eqn{S}, so that the right
 hand side of the first of equations \eqn{eqmot} will be proportional to
$\theta_\mu^a = \frac{\del {\cal L}}{\del e^\mu_a}~,~~$ which is
strictly related to the energy--momentum tensor, while the second
equation is unchanged. Then   the right hand side  of \eqn{eqm} is
proportional to  the energy momentum tensor $T_{\mu\nu}={\tt
g}_{\rho\nu} e^\rho_a \theta_\mu^a$. 

\section{General Relativity as a $q$--Gauge Theory}
In this section we will show that the equivalence between  Poincar\'e
gauge theory and Einstein gravity can be extended  to the case  where  
the local invariance with respect to the Lie algebra of $ISO(3,1)$ is 
replaced   by local invariance with respect to the {\it quantum Lie 
algebra} associated to the quantum Poincar\'e group $ISO_q(3,1)$. 

Let us first recall the definition of a quantum Lie algebra and its
connection to differential calculus on quantum groups, as described in
\cite{ac}. Starting from the definition of a quantum group $G_q$ as the
non--commutative algebra of functions on the Lie group $G$, $G_q\equiv
Fun_q(G)$, a bimodule of left (right) invariant forms for $G_q$ is
constructed, in the same way as the bimodule of left (right) invariant
forms is constructed for classical Lie groups. Such a bimodule inherits
the non--commutative nature of the product in $Fun_q(G)$, 
\be
R^{ab}_{ef} \, N^e_c N^f_d = N^b_f  N^a_e \,
R^{ef}_{cd}~,
\label{rtt}
\ee
($N$ is an element of $G$ in its defining representation, while $R$ is
the $R$--matrix, satisfying the quantum Yang Baxter equation) so that
the usual definition of exterior product for one--forms, $\theta^i
\wedge \theta^j =  \theta^i \otimes \theta^j - \theta^j
\otimes\theta^i$, is replaced on $q$--groups by 
\be
\theta^i \wedge \theta^j = \theta^i \otimes \theta^j -
\Lambda^{ij}_{kl} \, \theta^k \otimes \theta^l
\label{formprod}
\ee
where $\Lambda$ is the braiding matrix. Following the analogy with the
differential calculus on classical Lie groups, the algebra of left
invariant vector fields, which  is dual to the algebra of left invariant
one-forms, can be obtained,\footnote{In the same way we can introduce
right invariant objects. Bicovariant differential calculus \cite{woro}
requires  that left and right actions of the $q$--group on the bimodule
commute.} with $q$--commutation relations 
\be
T_i T_j - \Lambda_{ij}^{kl} \, T_k T_l \equiv [T_i, T_j]_q =C_{ij}^k T_k ~.
\label{qcom}
\ee
It is this algebra which is called the quantum Lie algebra. In the limit
$q\rightarrow 1$, $\Lambda^{kl}_{ij} \rightarrow \delta^k_j \delta^l_i$
and $T_i$ become the generators of the classical Lie algebra. 
$C_{ij}^k$ are $q$-structure constants, which in general are not
antisymmetric in the lower two indices except in the limit $q\rightarrow
1$. In order to define a bicovariant calculus, the braiding matrix
$\Lambda$ and the structure constants have to satisfy the following
relations \cite{cast1}: 
\beqa
\Lambda^{ij}_{kl}\Lambda^{lm}_{sp} \Lambda^{ks}_{qu} =
\Lambda^{jm}_{kl}\Lambda^{ik}_{qs} \Lambda^{sl}_{up}~~~
({\mbox{Yang~Baxter~
equation}}) \label{yb} \\
C^r_{mi} C^n_{rj}  -\Lambda^{kl}_{ij} C^r_{mk} C^n_{rl} =
C_{ij}^k C_{mk}^n~~~({\mbox{$q$-Jacobi}}) \label{qjac} \\
\Lambda^{ir}_{mk}\Lambda^{ks}_{nl} C_{rs}^j = \Lambda^{ij}_{kl} C_{mn}^k
~,
\label{2d} \\
\Lambda^{jq}_{ri}\Lambda^{si}_{kl} C_{ps}^r + \Lambda^{jq}_{pi} C_{kl}^i=
 C_{is}^j \Lambda^{sq}_{rl}\Lambda^{ir}_{pk} + C_{rl}^q\Lambda^{jr}_{pk}~.
\eeqa
The first condition is the quantum Yang Baxter equation; the second is
the Jacobi identity for the algebra \eqn{qcom}, while the last equations
are trivial in the limit $q\rightarrow 1$. 

Following \cite{cast1}, the gauge potential is assumed to be a $q$--Lie
algebra valued one--form $A \equiv A^i_\mu T_i dx^\mu$. In this approach
the deformation occurs solely in the fiber and thus the $A^i_{\mu}$ are
taken to be $q$--fields subject to nontrivial commutation relations.
Space--time, instead, remains an ordinary manifold so that $d x^{\mu}$
are ordinary space--time differentials commuting with $A^i_{\mu}$. The
exterior product of one--forms on the space--time  manifold is deformed
in the same way as the exterior product of invariant forms on the group
manifold \eqn{formprod} and, for general groups, one has: 
\be
A^i \wedge A^j = - Z^{ij}_{kl} \, A^{k} \wedge A^l~;  \label{com}
\ee
where $Z$ is a matrix of ordinary $c-$numbers which depends on the
group. The undeformed case obviously corresponds to the choice
$Z^{ij}_{kl}=\delta^i_l \delta^j_k$ for any group. Deformed Chern Simons
theories have been constructed only for minimal deformations
\cite{noiPL}, that is deformations satisfying $\Lambda^2= 1$; moreover,
only minimal deformations are known for the inhomogenous groups like
$ISO(3,1)$ which we are interested in \cite{cast96}. In this case the
matrix $Z$ has a simple expression in terms of the braiding matrix
$\Lambda$: 
\be
A^i \wedge A^j = - \Lambda^{ij}_{kl}\,A^{k} \wedge A^l~,  \label{qaa}
\ee
The braiding matrix $\Lambda$ will depend in general on a set of
parameters $q_{i}$ and on $r$. The number of independent parameters
depends on the group. For $ISO_q(3,1)$ there si only one parameter,
which we will indicate as $q$. The deformed gauge transformations are
assumed to be of the usual form 
\be
\delta_{\epsilon} A= -d\epsilon -A\epsilon + \epsilon A \label{var1}
\ee
where $\epsilon \equiv \epsilon^i T_i$. The gauge parameters
$\epsilon^i$ are now $q$--numbers and are assumed to have the following
commutation rules with the gauge fields: 
\be
\epsilon^i A^j = \Lambda^{ij}_{mn} A^m \epsilon^n~. \label{com1}
\ee
The commutation relations for $A^i$ with $d\epsilon^j$ and $dA^i$ with
$\epsilon^j$ can be obtained by taking the exterior derivative of the
above equation and imposing that the terms containing $dA^i$ and
$\epsilon^j$ cancel separately. The field strength is defined in the
usual way 
\be
F \equiv \frac{1}{2} F_{\mu \nu} dx^{\mu} dx^{\nu}
 = d A + A ^2~,   \label{fistrength}
\ee
where $A^2 = A^i A^j T_i T_j$. $F$ is valued in the deformed Lie-algebra
\cite{cast1} and under a gauge transformation \eqn{var1} it transforms
as: 
\be
\delta_{\epsilon} F = \epsilon F - F \epsilon~.\label{var2}
\ee
Let us specialize now to the $q$--Poincar\'e group. The group manifold
is parameterized by a Lorentz vector $z_a$ and a Lorentz matrix
$\ell_{ab}$ so that \eqn{rtt} is replaced by 
\be
z^a ~{\ell_c}^{b} = q^{\Delta(b)} ~ {\ell_c}^{b} ~z^a ~,\label{4dcz}
\ee
where
\be
\Delta(1)=-1 ~,\quad \Delta(2)=\Delta(3)=0 ~,\quad \Delta(4)=1 ~,
\label{Delts}
\ee
and all other commutation relations are trivial. The Lorentz metric
tensor is taken to be the following off-diagonal matrix: 
\be
\eta=\pmatrix{ & & & 1\cr & 1 & & \cr & & 1 & \cr 1 & & & \cr}~.
\label{eta}
\ee
Then the  commutation relations \eqn{4dcz}  are consistent with the
Lorentz constraints \eqn{dolm} due to the identity
\be
\eta_{ab}= q^{\Delta(a)+\Delta(b)} \eta_{ab} ~,
\ee
$ISO_q(3,1)$ thus contains the undeformed Lorentz group. The braiding
matrix, $\Lambda$, appearing in \eqn{qcom} and \eqn{qaa} is given by 
\beqa
\Lambda_{ab~cd}^{cd~ab}=1 ~,\quad
\Lambda_{a~b}^{b~a}= q^{\Delta(a)- \Delta(b)} ~, \cr
\Lambda_{bc~a}^{a~bc}= (\Lambda_{a~~bc}^{bc~~a})^{-1}=
q^{\Delta(b)+\Delta(c)} ~,
\eeqa
with all other components vanishing. In terms of the Lorentz and
translation generators, $M_{ab}$ and $P_a$, the quantum Lie algebra
\eqn{qcom} is expanded to 
\beqa
[M_{ab}, M_{cd}] = \eta_{ac} M_{bd} -&& \eta_{bc} M_{ad} +
\eta_{bd} M_{ac} - \eta_{ad} M_{bc}\cr
[M_{ab}, P_c]_{q^{\Delta (a)+ \Delta (b)}} &&= -(\eta_{bc} P_{a}
- \eta_{ac} P_{b} )\cr
[P_a, P_b]_{q^{\Delta (a)- \Delta (b)}} &&= 0~, \label{qlie}
 \eeqa
where $[\alpha,\beta]_s\equiv\alpha\beta - s\beta\alpha$. Eqs.
\eqn{qlie} reduce to \eqn{lie} for $q=1$.  For each value of $q$ the
quantum Lie-algebra  contains the undeformed Lorentz algebra. 

Splitting the connection one--form into spin--connection, 
$
~\omega(x) =\omega_\mu^{ab}(x)M_{ab} dx^\mu ~
$
and vierbein one--form, 
$~e(x)=e_\mu^a(x)P_a dx^\mu~,~ $ 
as in the undeformed case, and assuming the space--time to be spanned by
ordinary commutative coordinates, \eqn{qaa} implies 
\beqa
\omega _\mu ^{ab}\omega _\nu ^{cd}=\omega _\nu ^{cd}\omega _\mu 
^{ab}~,\cr
e_\mu ^a\omega _\nu ^{bc}=q^{\Delta (b)+\Delta (c)}~
\omega _\nu ^{bc}e_\mu^a~,\cr
e_\mu ^ae_\nu ^b=
q^{\Delta (b)-\Delta (a)}~ e_\nu ^be_\mu ^a~.
\label{4dcrstc}
\eeqa
Splitting the gauge parameter $\epsilon^i$ into Lorentz and translation
parameter, $\tau, \rho$ respectively, the gauge transformations \eqn{var1} 
become 
\begin{eqnarray}
\delta \omega^{ab} & =& d\tau^{ab} + {\omega^a}_c \;\tau^{cb} -
{\omega^b}_c \;\tau^{ca} \;,\cr
\delta e^c & = & d\rho^c +
{\omega^{c}}_b \;\rho^b - \tau^{cb}\;e_b\;.  \label{4dtras}
\end{eqnarray}
From \eqn{com1} we get the following commutation relations between 
gauge parameters and one--forms 
\begin{eqnarray}
\rho^a \;\omega^{bc} & =& q^{\Delta(b)+\Delta(c) } \;\omega^{bc}\;
\rho^a \cr
\rho^a\; e^{b} & =& q^{\Delta(b)-\Delta(a)} \; e^{b}\;  \rho^a \cr
\tau^{ab} \; e^c & =& q^{-\Delta(a)-\Delta(b)} \; e^{c} \;
\tau^{ab}\;~ \cr
\tau^ {ab} \omega^{cd} &=& \omega^{cd}  \tau^ {ab} \;. \label{crgpc}
\end{eqnarray}
Finally the curvature and the torsion ${\cal R}^{ab}$ and ${\cal T}^a $,
have the usual expressions 
\begin{eqnarray}
{\cal R}^{ab} & =& d\omega^{ab}+ {\omega^a}_c\wedge \omega^{cb}\;, \cr
{\cal T}^a & =& de^a+ {\omega^a}_b\wedge e^b\;,  \label{4dRT}
\end{eqnarray}
though they obey non trivial braiding relations with the connection 
components, which can be obtained by \eqn{4dcrstc}.

Next we write down  a locally Lorentz invariant action:
\be
{\cal S} =\frac{1}{4} \int_M \; \epsilon_{abcd} {\cal R}^{ab} \wedge {\cal
E}^{cd}
\;,
\label{4dact}
\ee
where ${\cal E}^{cd} $ is the two--form
\be
{\cal E}^{cd}= -{\cal E}^{dc} = q^{-\Delta(d)} e^c \wedge e^d \;,
\ee
$M$ is a four manifold and $\epsilon_{abcd} $ is the ordinary, totally
antisymmetric tensor with $\epsilon_{1234} =1$. The expression
(\ref{4dact}) differs from that of  the undeformed case by the
$q^{-\Delta(d)} $ factor. Note that this factor can be written
differently using the identity 
\be
q^{\Delta(a)+\Delta(b)+ \Delta(c)+ \Delta(d)} \; \epsilon_{abcd}
=\epsilon_{abcd}\;\;.  \label{epsid}
\ee
As in the undeformed case, the action is invariant  under the full set
of local Poincar\'e  transformations (\ref{4dtras}), provided we impose
the torsion to be zero upon making the variations. 

The equations of motion obtained from varying the vierbeins have the
usual form, i.e. 
\be
\; \epsilon_{abcd} {\cal R}^{ab} \wedge e^{c} = 0 \;,  \label{ffeq1}
\ee
while varying $\omega^{ab}$ gives
\be
d\tilde {\cal E}_{ab} = {\omega_a}^c \tilde {\cal E}_{bc} - {\omega_b}^c
\tilde {\cal E}_{ac}
\;,\qquad \tilde {\cal E}_{ab}\equiv \epsilon_{abcd} {\cal E}^{cd} \;.
 \label{eomfcE}
\ee
Due to the antisymmetry of ${\cal E}^{cd}$, we get the following
expression in terms of the torsion form  (\ref{eomfcE}) 
\be
\; \epsilon_{abcd} {\cal T}^{c} \wedge e^{d}\; q^{-\Delta(d)} = 0 \;.
\label{ffeq2}
\ee
We will show next that this equation implies zero torsion, provided
inverse vierbeins exist. This is necessary in order to recover
Einstein's gravity. 

\section{Recovering Einstein's theory}
We now prove that the metric formulation of the q-deformed Cartan theory
of gravity discussed above is completely equivalent to the {\it
undeformed} Einstein's theory, for all values of $q$. 

To make a connection with Einstein gravity, we need to introduce the
space-time metric ${\tt g}_{\mu \nu}$ on $M$. As in the undeformed case
it has to be a bilinear in  the vierbeins which is symmetric in the
space-time indices and invariant under local Lorentz transformations.
These requirements uniquely fix ${\tt g}_{\mu\nu}$ to be 
\be
{\tt g}_{\mu \nu }=q^{\Delta (a)}\;\eta _{ab}\;e_\mu ^ae_\nu ^b\;,
\label{4dsymmet}
\ee
Using eqs.\eqn{4dcrstc} we see that ${\tt g}_{\mu \nu }$ is symmetric,
although the tensor elements are not c-numbers since 
\beqa
{\tt g}_{\mu \nu }\;\omega _\rho ^{ab}=q^{2\Delta (a)+2\Delta (b)}\;
\omega_\rho ^{ab}\;{\tt g}_{\mu \nu }\;,\cr {\tt g}_{\mu \nu }\;e_\rho
^a=q^{2\Delta (a)}\;e_\rho ^a\;{\tt g}_{\mu \nu }\;.
\eeqa
The components of ${\tt g}_{\mu \nu }$ do however commute with themselves.

The inverse $e^{\mu}_a$ of the vierbeins $e^a_{\mu}$ can be defined 
if we enlarge our algebra by a new element ${\tt e}^{-1}$ such that: 
\beqa
{\tt e}^{-1} e^a_{\mu}&=& q^{-4 \Delta(a)}e^a_{\mu} \; {\tt e}^{-1}~,\\
{\tt e}^{-1} \omega^{ab}_{\mu}&=& q^{-4 (\Delta(a)+ \Delta(b))
}\omega^{ab}_{\mu}\; {\tt e}^{-1}~\label{creminus}\\
{\tt e}^{-1} {\tt e} &=&1~,\label{eeminus}
\eeqa
where ${\tt e}$ is the determinant:
\be
{\tt e}=\epsilon^{\mu \nu \rho \sigma} e^1_{\mu} e^2_{\nu} e^3_{\rho}
e^4_{\sigma}~.
\ee
Eq.\eqn{eeminus} is consistent because its left hand side  commutes with
everything, due to eqs.\eqn{creminus}. Moreover, one can check that
${\tt e}^{-1} {\tt e} = {\tt e } {\tt e}^{-1}$. The inverse of the
vierbeins can now be written: 
\be
e^{\mu}_a= \frac{1}{3!} \hat{\epsilon}_{abcd} \epsilon^{\mu \nu \rho
\sigma} e^b_{\nu} e^c_{\rho} e^d_{\sigma} {\tt e}^{-1}~, \label{definv}
\ee
where the totally q-antisymmetric tensor  $\hat{\epsilon}_{abcd} $  is
defined such that 
\be
\hat{\epsilon}_{abcd} e^a \wedge e^b \wedge e^c \wedge e^d = e^1 \wedge e^2
\wedge e^3 \wedge e^4 ~~~{\rm no~sum~on}~a,b,c,d
\ee
The solution to this equation can be expressed by
\be
\hat{\epsilon}_{abcd} = q^{3\Delta(a)+2\Delta(b)+\Delta(c) +3}\;
\epsilon_{abcd}  \;.\label{ephaep}\end{equation}
Notice also the following useful identity satisfied by the q-antisymmetric
tensor $\hat{\epsilon}^{abcd}$ obtained by raising the indices of
$\hat{\epsilon}_{abcd}$ with the metric $\eta^{ab}$
\be q^{-6}\; \hat{\epsilon}^{abcd}\;{\tt e}
=-\epsilon^{\mu\nu\lambda\sigma} e^a_\mu e^b_\nu e^c_\lambda e^d_\sigma
\;.\label{epep} 
\ee
The explicit expression of $\hat{\epsilon}^{abcd}$ can be seen to
be:
\be
\hat{\epsilon}^{abcd} = q^{-3\Delta(a)-2\Delta(b)-\Delta(c) +3}\;
\epsilon^{abcd}  \;,\label{defeup} 
\ee
where $\epsilon^{abcd}$ is the ordinary antisymmetric tensor obtained by
raising the indices of $\epsilon_{abcd}$ with the metric $\eta^{ab}$. It is
easy to prove that the inverse of the vierbeins
\eqn{definv} have the usual properties:
$$
e^a_{\mu} e_b^{\mu}=e_b^{\mu} e^a_{\mu} = \delta^a_b~,
$$
\be
e^a_{\mu} e_a^{\nu} = e_a^{\nu} e^a_{\mu} = \delta^{\nu}_{\mu}~.
\label{prode}
\ee

By using the inverse of the vierbeins, we can now prove that
eq.\eqn{ffeq2} implies the vanishing of the torsion. To begin with, we
introduce the components of the torsion two-form along the vierbeins: 
\be
{\cal T}^a_{bc} \; \equiv \; q^{\Delta(b)} {\cal T}^a_{\mu \nu} e^{\mu}_b
e^{\nu}_c~; 
\ee
the power of $q$ ensures that they are antisymmetric in the lower
indices, ${\cal T}^a_{bc}=-{\cal T}^a_{cb}$. Now we rewrite
eq.\eqn{ffeq2} as: 
\beqa
0&=&q^{-\Delta(d)}\; \epsilon_{abcd} \epsilon^{\lambda \mu \nu
\rho} {\cal T}^c_{\mu \nu}e^d_{\rho}=
q^{-\Delta(d)-\Delta(h)}\; \epsilon_{abcd} \epsilon^{\lambda \mu \nu
\rho} {\cal T}^c_{gh} e^g_{\mu} e^h_{\nu} e^d_{\rho}=\nonumber\\
&=& - q^{\Delta(d) - \Delta(f)-3} \epsilon_{abcd}
\epsilon^{fghd} {\cal T}^c_{gh} e^{\lambda}_f {\tt e}=\nonumber\\
&=&2 q^{- \Delta(a) -\Delta(b) -\Delta(c) -3}(q^{-\Delta(a)}{\cal T}^c_{bc}
e^{\lambda}_a + q^{-\Delta(b)}{\cal T}^c_{ca} e^{\lambda}_b +
q^{-\Delta(c)}{\cal T}^c_{ab} e^{\lambda}_c)\;{\tt e}~,\label{torzero}
\eeqa
where we have used the identity
\be
q^{-\Delta(d)-\Delta(h)}\epsilon^{\lambda \mu \nu \rho}e^g_{\mu} e^h_{\nu}
e^d_{\rho}= -q^{\Delta(d)-\Delta(f)-3} \epsilon^{fghd} e^{\lambda}_f {\tt
e}~,
\ee
which follows from eqs.\eqn{epep} and \eqn{defeup}. Neglecting the overall
factor of $q^{-\Delta(a)
-\Delta(b)
-3}{\tt e}$ in
\eqn{torzero} and multiplying it on the right by $e^d_{\lambda}$ we finally
get
\be
q^{-\Delta(c)}{\cal T}^c_{bc} \; \delta^d_a + q^{-\Delta(c)}{\cal T}^c_{ca}
\; \delta^d_b +q^{- \Delta(d)} {\cal T}^d_{ab}=0~.
\ee
It is easy to verify that these equations imply the vanishing of all the
${\cal T}^a_{bc}$ and thus of the torsion.

The Christoffel symbols $\Gamma _{\mu \nu }^\sigma $ are defined in the 
same way of the previous section, by
demanding that the covariant derivative of the vierbeins vanishes, 
\be
{\tt D}_\mu e_\nu ^b=0~.\label{coveq}
\ee 
The difference with the undeformed case is that we cannot switch the
order of objects arbitrarily. To eliminate the spin--connection from
\eqn{covd} we now multiply on the left by $q^{\Delta (a)}\eta
_{ab}e_\rho ^a$,  and proceed as in the undeformed case. We can then
isolate $\Gamma _{\mu \nu}^\sigma $ according to 
\be
2q^{\Delta (a)}\eta _{ab}e_\rho ^ae_\sigma ^b\Gamma _{\mu \nu }^\sigma
=q^{\Delta (a)}\eta _{ab}[e_\rho ^a(\partial _\mu e_\nu ^b+\partial _\nu
e_\mu ^b)+e_\nu ^a(\partial _\mu e_\rho ^b-\partial _\rho e_\mu ^b)+e_\mu
^a(\partial _\nu e_\rho ^b-\partial _\rho e_\nu ^b)]
\ee
or
\be
2{\tt g}_{\rho \sigma }\Gamma _{\mu \nu }^\sigma =\partial _\mu {\tt g}
_{\rho \nu }+\partial _\nu {\tt g}_{\rho \mu }-\partial _\rho {\tt g}_{\nu
\mu }\;.  \label{qchris}
\ee
To solve this equation we need the inverse of the metric ${\tt g}^{\mu
\nu}$. The expression 
\be
{\tt g}^{\mu \nu}= q^{\Delta(a)} \eta^{ab} e^{\mu}_a e^{\nu}_b  \;,
\ee
does the job as it can be checked that
\be
{\tt g}^{\mu \rho} {\tt g}_{\rho \nu}=
{\tt g}_{\nu \rho} {\tt g}^{\rho \mu}= \delta^{\mu}_{\nu} \;.
\ee
Notice that  unlike in the usual Einstein Cartan theory
\be
{\tt g}^{\mu \nu} \eta_{ab} e^b_{\nu}= q^{\Delta(a)} e^{\mu}_a~.\label{unl}
\ee
We are now able to solve eq.\eqn{qchris}. Upon multiplying it by ${\tt
g}^{\tau \rho}$ on each side, we get the usual expression for the
Christoffel symbols in terms of the metric tensor and its inverse. It
may be verified, using these expressions, that the Christoffel symbols
commute with everything and thus, even if written in  terms of
non-commuting quantities, they can be interpreted as being ordinary
numbers. 

The covariant derivative operator $\nabla_{\mu}$ defined by the
Christoffel symbols is compatible with the metric ${\tt g}_{\mu \nu}$,
i.e.  $\nabla_{\mu} {\tt g}_{\nu \rho}=0$. This is clear because our
Christoffel symbols have the standard expression in terms of the
space-time metric ${\tt g}_{\mu \nu}$, but also follows from
eq.\eqn{coveq} 
\be
\nabla_{\mu} {\tt g}_{\nu \rho}={\tt D}_{\mu} {\tt g}_{\nu \rho}=
{\tt D}_{\mu}( q^{\Delta (a)} \eta_{ab} e^a_{\nu} e^b_{\rho})=0~.
\ee
We now construct the Riemann tensor. It is defined as in the undeformed
theory:
\be
{\tt R}_{\mu \nu \rho}^{~~~~\sigma} v_{\sigma} =({\tt D}_{\mu} {\tt
D}_{\nu}-{\tt D}_{\nu} {\tt D}_{\mu} ) v_{\rho}~,
\ee
where $v_{\mu}$ is an ordinary co-vector. It follows from \eqn{coveq}
that it has the standard expression in terms of the Christoffel symbols
(and thus in terms of the space-time metric and its inverse) and
therefore its components commute with everything. (This is also true for
the Ricci tensor ${\tt R}_{\mu \nu}={\tt R}_{\mu \sigma
\nu}^{~~~~\sigma}$, of course, but not for ${\tt R}_{\mu \nu \rho \tau}$
as the lowering of the upper index of the Riemann tensor implies
contraction with ${\tt g}_{\sigma \tau}$ which is not in the center of
the algebra). The relation among the Riemann tensor and the curvature of
the spin connection follows from eq. \eqn{coveq}: 
\be
e_{\sigma}^a {\tt R}_{\mu \nu \rho}^{~~~~\sigma} v^{\rho}
=e_{\sigma}^a({\tt D}_{\nu} {\tt D}_{\mu}- {\tt D}_{\mu} {\tt D}_{\nu})
 v^{\sigma}=({\cal
D}_{\nu} {\cal D}_{\mu}- {\cal D}_{\mu} {\cal D}_{\nu})e_{\sigma}^a
v^{\sigma}= -{\cal R}_{\mu
\nu }^{ac}\eta_{bc} e_{\sigma}^b v^{\sigma}~,
\ee
$ {\cal R}^{ab}_{\mu \nu}$ being the space-time components of $ {\cal
R}^{ab}$. $v^{\mu}$ being an arbitrary ordinary vector, it follows from
the above equation that: 
\be
{\tt R}_{\mu \nu \rho}^{~~~~\tau}=-{\cal R}^{ac}_{\mu \nu }\eta_{bc}
e^b_{\rho}e^{\tau}_a~.\label{riecur}
\ee
Using this equation it can be checked directly that the components of
the Riemann tensor commute with everything, as pointed out earlier. Our
Riemann tensor has the usual symmetry properties: 
\beqa
{\tt R}_{\mu \nu \rho}^{~~~~\sigma}& = & -{\tt R}_{\nu \mu
\rho}^{~~~~\sigma}~,\nonumber\\{\tt R}_{\mu \nu
\rho \sigma}& =& - {\tt R}_{\mu
\nu \sigma \rho}
~,\nonumber\\{\tt R}_{[\mu \nu \rho]}^{~~~~~\sigma}& = & 0~. \label{riem}
\eeqa
The first of these equations is obvious; the second can be proved
starting from \eqn{riecur}: 
\beqa
&{\tt R}_{\mu \nu \rho \sigma}&={\tt R}_{\mu \nu \rho}^{~~~~\tau} {\tt
g}_{\tau
\sigma}=
-{\cal R} ^{ab}_{\mu \nu} e_{\rho b}e^{\tau}_a 
{\tt g}_{\tau \sigma}=\nonumber\\
&=&-q^{\Delta(a)}{\cal R} ^{ab}_{\mu \nu}e_{\rho b} e_{\sigma a} =
-q^{\Delta(b)}{\cal R} ^{ab}_{\mu \nu}e_{\sigma a} e_{\rho b}=
-{\tt R}_{\mu \nu \sigma \rho}~,
\eeqa
where we have made use of \eqn{unl}. The third of eqs.\eqn{riem} follows
from the algebraic Bianchi identity and from \eqn{riecur}: 
\be
0=-\epsilon^{\lambda \mu \nu \rho}{\cal R}^{ac}_{\mu \nu } \eta_{bc}
e^b_{\rho}=\epsilon^{\lambda \mu \nu \rho}{{\tt R}_{\mu \nu
\sigma}}^{\tau} e^a_{\tau} e^{\sigma}_b e^b_{\rho}=6 \;
\epsilon^{\lambda \mu \nu
\rho}{\tt R}_{[\mu
\nu \rho]}^{~~~~\tau} e^a_{\tau}~.
\ee

We now show that the action \eqn{4dact} becomes equal to the {\it
undeformed} Einstein-Hilbert action, once the spin connection is
eliminated using its equations of motion, namely the zero torsion
condition. As in the undeformed case, first we rewrite \eqn{4dact} in a
form analogous to Palatini's action, and then show that the latter
reduces to the {\it undeformed} Einstein-Hilbert action, once the
spin-connection is eliminated from it. Consider thus the following
deformation of the Palatini action: 
\be
{\cal S}=\frac{1}{2}\int_M d^4 x \, q^{\Delta(a)-3} {\tt e} \, e^{\mu}_a
e^{\nu}_b {\cal R}^{ab}_{\mu \nu}~.
\label{qpala}
\ee
To see that it coincides with \eqn{4dact}, we use the identity: 
\be 
q^{\Delta(a)-\Delta(b)-6}    \hat{\epsilon}^{abcd}
 e^\mu_a e^\nu_b {\tt e}   =
-\epsilon^{\mu\nu\lambda\sigma} e^c_\lambda e^d_\sigma \;. \label{id68}
\ee  
The result (\ref{qpala}) then follows after multiplying both sides of
this equation on the left by 
$
-1/8\; q^{-2\Delta(f) - \Delta(g)-3}
\hat{\epsilon}_{fgcd} {\cal R}^{fg}_{\mu\nu} 
$ 
and using the identity
$$
\hat{\epsilon}_{fgcd}
 \hat{\epsilon}^{abcd} =-2 q^6\;(\delta^a_f\delta^b_g-q^{\Delta(f)-\Delta(g)}
\delta^a_g\delta^b_f)\;,
$$ 
along with (\ref{ephaep}).

We now show that eq.\eqn{qpala} becomes in turn equal to the undeformed
Einstein-Hilbert action upon eliminating the spin connection via its
equation of motion. This amounts to expressing ${\cal R}^{ab}_{\mu \nu}$ in
terms of the Riemann tensor by inverting eq.\eqn{riecur} and then plugging
the result in eq.\eqn{qpala}. We have:
\beqa
&&q^{\Delta(a)} e^{\mu}_a e^{\nu}_b {\cal R}^{ab}_{\mu \nu}=-q^{\Delta(a)}
{\tt R}_{\mu \nu \rho}^{~~~~\tau} e^{\mu}_a e^{\nu}_b e^a_{\tau}
e^{b\rho}=\nonumber\\ &&=-q^{\Delta(b)} {\tt R}_{\mu \nu
\rho}^{~~~~\mu}e^{\nu}_b e^{b\rho}= {\tt R}_{\nu \mu \rho}^{~~~~\mu} {\tt
 g}^{\nu
\rho}={\tt R}~,\label{einhil1}
\eeqa
where we have made use of \eqn{prode}. Moreover we get, after a 
cumbersome calculation:
\be
{\tt g} \equiv \det \parallel {\tt g}_{\mu \nu} \parallel =
\frac{1}{4!}\epsilon^{\mu_1 \mu_2 \mu_3
\mu_4} \epsilon^{\nu_1 \nu_2 \nu_3 \nu_4} {\tt g}_{\mu_1 \nu_1}
 {\tt g}_{\mu_2 \nu_2} {\tt g}_{\mu_3 \nu_3} {\tt g}_{\mu_4 \nu_4}
 =q^{-6} \; {\tt
 e}^2~,\label{einhil2}
 \ee
Putting together \eqn{einhil1} and \eqn{einhil2} we see that the
q-Palatini action \eqn{qpala} becomes equal to: 
\be
{\cal S}=\frac{1}{2}\int_M d^4 x \, \sqrt{-{\tt g}}\; {\tt R}~,
\ee
which is the {\it undeformed} Einstein-Hilbert action. Since the components
of ${\tt g}_{\mu \nu}$ and its inverse all commute among themselves, it is
clear that the equations of motion of the metric theory will be equal
to those of the undeformed Einstein's theory in vacuum. One can obtain the
same result starting directly from eq.\eqn{ffeq1} and using \eqn{riecur}.

\section{Conclusions}

From the results of the last section we may conclude that if we just
consider the  theory constructed in terms of the space-time metric ${\tt
g}_{\mu \nu}$, ignoring the underlying gauge formulation,  our theory is
completely equivalent to Einstein's theory.  No trace of the
non--commutative structure existing in the gauge formulation of the
theory can be found at the metric level. Though the metric itself is
non--commutative, as it doesn't commute with the connection components,
all the physical objects constructed out of it, e. g. the Christoffel
symbols together with the Riemann, Ricci and Einstein tensors, are
c-number. Thus it appears that, at the level of {\it classical} General
Relativity we can choose whatever representative of the one parameter
family of $q$--gauge theories (not only the well known $q=1$ theory)
without changing the physics we are describing. That is, we have
discovered a non--commutative structure in General Relativity which is
hidden, even in the presence of matter, provided there are no sources of
torsion. The possible physical implications of such an hidden structure
need however further, careful investigation. We will report on these
aspects in a forthcoming paper. 

\section{Acknowledgments}
P.V. would like to thank the Organizing Committee   
of Quantum Groups-97, and especially M. Arik, for inviting this contribution.

\end{document}